\documentclass[traditabstract]{aa}
\usepackage{graphicx}
\usepackage{txfonts}

\usepackage{natbib}
\usepackage{color}
\usepackage{overpic}
\newcommand{\teff}{$T_{\rm{eff}}$}
\newcommand{\logg}{$\log{g}$}
\newcommand{\feh}{[Fe/H]}

\begin{document} 

   \title{Inelastic e+Mg collision data and its impact on modelling stellar and supernova spectra \thanks{Full tables 2-3 are 
   only available in electronic form at the CDS via anonymous ftp to cdsarc.u-strasbg.fr (130.79.128.5) 
   or via http://cdsweb.u-strasbg.fr/cgi-bin/qcat?J/A+A/.}}
   
   \titlerunning{Inelastic e+Mg collisions}

   %\subtitle{}

\author{P. S. Barklem\inst{1}
          \and
        Y. Osorio\inst{2,3,4}  
          \and 
        D. V. Fursa\inst{5}
          \and
        I. Bray\inst{5}
          \and 
        O. Zatsarinny\inst{6}
          \and
        K. Bartschat\inst{6}
          \and
        A. Jerkstrand\inst{7}
          }

   \institute{Theoretical Astrophysics, Department of Physics and Astronomy, Uppsala University,
              Box 516, SE-751 20 Uppsala, Sweden  
         \and
              Instituto de Astrof\'isica de Canarias, v\'ia L\'actea, 38205 La Laguna, Tenerife, Spain
         \and
              Departamento de Astrof\'isica, Universidad de La Laguna, Tenerife, Spain
         \and
              Max-Planck-Institut f\"ur Astronomie, K\"onigstuhl 17, 69117, Heidelberg, Germany
         \and
              Curtin Institute for Computation and Department of Physics, Astronomy and Medical Radiation Science, 
              Kent Street, Bentley, Perth, Western Australia 6102, Australia
         \and     
              Department of Physics and Astronomy, Drake University, Des Moines, IA 50311, USA
         \and
              Max-Planck Institut f\"ur Astrophysik, Karl-Schwarzschild-Str. 1, D-85748 Garching, Germany
             }

   \date{Received soon ; accepted later}

% \abstract{}{}{}{}{} 
% 5 {} token are mandatory
 
  \abstract
   {Results of calculations for inelastic e+Mg effective collision strengths for the lowest 25 
   physical states of \ion{Mg}{i} (up to 3s6p$\ ^1$P), and thus 300 transitions, from the convergent 
   close-coupling (CCC) and the $B$-spline $R$-matrix (BSR) methods are presented.  At temperatures 
   of interest, $\sim$ 5000~K, the results of the two calculations differ on average by only 4\%, 
   with a scatter of 27\%.  As the methods are independent, this suggests that the calculations 
   provide datasets for e+Mg collisions accurate to this level.  Comparison with the 
   commonly used dataset compiled by \cite{Mauas1988}, covering 25 transitions among 12 states, suggests 
   the \citeauthor{Mauas1988} data are on average $\sim$57\% too low, and with a very 
   large scatter of a factor of $\sim$6.5.  In particular the collision strength for 
   the transition corresponding to the \ion{Mg}{I} intercombination line at 457~nm is significantly underestimated by \citeauthor{Mauas1988}, 
   which has consequences for models that employ this dataset.  In giant stars the new data leads to a stronger line compared to previous non-LTE calculations, and thus a reduction in the non-LTE abundance correction by $\sim$0.1 dex ($\sim$25~\%).  A non-LTE calculation in a supernova ejecta model shows this line becomes significantly stronger, by a factor of around two, alleviating the discrepancy where the 457~nm line in typical models with Mg/O ratios close to solar tended to be too weak compared to observations.}


   \maketitle
%
%-----------------------------------------------------------------

\section{Introduction}

Magnesium is an element of significant astrophysical importance.  It is an $\alpha$-element 
produced by hydrostatic carbon burning in massive stars
and ejected in core-collapse supernovae occurring just a few years after
carbon burning is complete (C burning itself takes a few hundred years).
Its significant abundance combined with its array of spectral features across the UV, 
visual, and IR spectrum, means that it is observable in stars of all spectral types, and even 
in the most metal-poor stars.  Its production depends sensitively on the progenitor mass of the
star \citep[e.g.][]{woosley_evolution_1995,limongi_massive_2000}, and is therefore an important diagnostic of
this in supernova spectral analysis \citep[e.g.][]{fransson_late_1989}.
From an atomic physics point of view, Mg is also attractive 
for study, as it is a quasi-two-electron system and thus often tractable to full-quantum 
treatments using recently-developed computational methods.   

Modelling of astrophysical gas and plasma out of equilibrium, i.e.\ in non-LTE, requires 
that all relevant radiative and collisional processes on the atom of interest be described, 
and modern high quality observations demand better modelling, which in turn demands better 
input atomic data.  For any astrophysical gas with a significant degree of ionisation, 
collisions involving electrons are very important, due to predominantly their higher 
speed than the heavier particles, which in turn leads to relatively large collision 
rates and usually larger cross sections.

Non-LTE modelling of Mg is done for a range of purposes.  Some recent examples include 
the measurement of stellar abundances in cool stars \citep{bergemann_red_2015, osorio_mg_2015, osorio_mg_2016} and in hot 
stars \citep{Przybilla2001}, the use of the \ion{Mg}{i} inter\-combination line at 457~nm as a temperature diagnostic 
of the solar chromosphere \citep{Langangen2009}, and as a probe of velocity fields in 
giants stars \citep{vieytes_modelling_2011}.   In addition, IR lines of Mg 
at 12 and 18~$\mu$m that are seen in emission in some cool stars \citep{Chang1983} may 
have future diagnostic potential, and they are known to form in far from LTE 
conditions \citep{Chang1991,Carlsson1992,Sundqvist2008,osorio_mg_2015}.   Mg lines are important in the modelling of spectra of supernovae 
ejecta \citep{jerkstrand_progenitor_2012,jerkstrand_late-time_2015}, where the main line seen in nebular SN spectra is the \ion{Mg}{i} 457~nm intercombination line, although \ion{Mg}{i} 1.50~$\mu$m is also normally seen when NIR observations are taken.

Many of the studies mentioned above use data for inelastic electron collision 
processes on neutral Mg assembled by \cite{Mauas1988}.  These data were collected 
from various sources, and are a mixture of theoretical, semi-empirical, and experimental results.  
Recently, attempts have been made to improve the situation and provide a coherent and consistent dataset 
from modern close-coupling calculations, namely the $B$-spline $R$-matrix (BSR) method 
and the $R$-matrix with pseudo\-states (RMPS) method.  \cite{merle_effective_2015} 
presented data based on the BSR method \citep{zatsarinny_cross_2009}.  However, this 
dataset did not include triplet-triplet transitions, and thus is of limited usefulness 
for modelling the Mg atom as a whole.  \cite{osorio_mg_2015} presented data obtained 
with the RMPS method.  In this case the data cover only the lowest 10 states, and for 
higher-lying states had to be complemented by the use of the impact parameter 
method \citep{Seaton1962b} for optically allowed transitions, and  in the case 
of forbidden transitions by assuming similar behaviour as the $R$-matrix calculations 
for low-lying states.  We note that various lines of interest, for example at 552.8~nm, 
involve such higher-lying states.  

Thus, though the recent work of \cite{merle_effective_2015} and \cite{osorio_mg_2015} 
are no doubt improvements on the data in \cite{Mauas1988}, both works have shortcomings 
in terms of the completeness one would wish for in non-LTE modelling of the entire Mg atom.  
The aim of the present work, therefore, is to provide a set of electron collision data for neutral magnesium 
that is both reliable and approaching the level of completeness required.  
For this purpose, calculations have been performed using the convergent close-coupling (CCC) 
method, as well as new calculations with the BSR method.  In this paper we will present 
effective collision strengths calculated from these methods for transitions between 
the 25 lowest-lying states (i.e. up to 3s6p$\ ^1$P).  The results are compared to 
give an indication of the reliabilities of the datasets.
   
%-----------------------------------------------------------------
\section{Theory and calculations}

Calculations have been done using the CCC \citep{bray_calculation_1995,bray_electrons_2002}
and BSR 
methods \citep[see][and references therein]{zatsarinny_b_2013}. 
These calculations are now detailed in turn.  

The CCC method has already been extensively applied to the electron-magnesium collision 
system \citep{fursa_excitation_2001,Bartschat2004,brown_electron_2005,bray_calculation_2015}. 
The atomic target is treated as two-electrons above an inert Hartree-Fock 
core~\citep{fursa_convergent_1997}. 
The target states are acquired by first obtaining one-electron orbitals by diagonalising 
the Mg$^+$ Hamiltonian in a Laguerre basis, and then constructing two-electron configurations 
from the orbitals and diagonalising the Mg Hamiltonian. The states are used to define 
the close-coupling equations in momentum space, which are solved separately for every 
partial wave and incident energy. The convergence of the results is tested by increasing 
the size of the Laguerre basis, and utilising the resulting increased number of states in the calculations. 
The present CCC results 
were obtained by taking the Laguerre basis size to be $N_l=30-l$ for
$l_{\rm max}=4$, which generated a total of 259 states. These were
checked for convergence against calculations with $N_l=25-l$, which
generated 209 states. Only the results of the largest CCC calculation
are being presented.
 
For the present work a new calculation has been performed with the BSR method, which improves 
upon the earlier calculation presented in \cite{zatsarinny_cross_2009} 
and \cite{merle_effective_2015}.  This model includes 32 bound states of 
Mg plus 680 continuum pseudo\-states with energies up to 30 eV.  These $R$-matrix 
with pseudo\-states calculations are similar to the recent BSR calculations 
for e+Be \citep{zatsarinny_calculations_2016} and e+B \citep{wang_electron-impact_2016}, 
where more computational details are given.  The target states in the BSR calculations 
were generated in the same scheme as in \cite{zatsarinny_cross_2009}; however, 
the R-matrix radius was reduced from 80 to 45 a.u.\ to cover a larger range 
in the Mg pseudo\-spectrum.  

\begin{figure}
   \centering
   \begin{overpic}[width=0.24\textwidth]{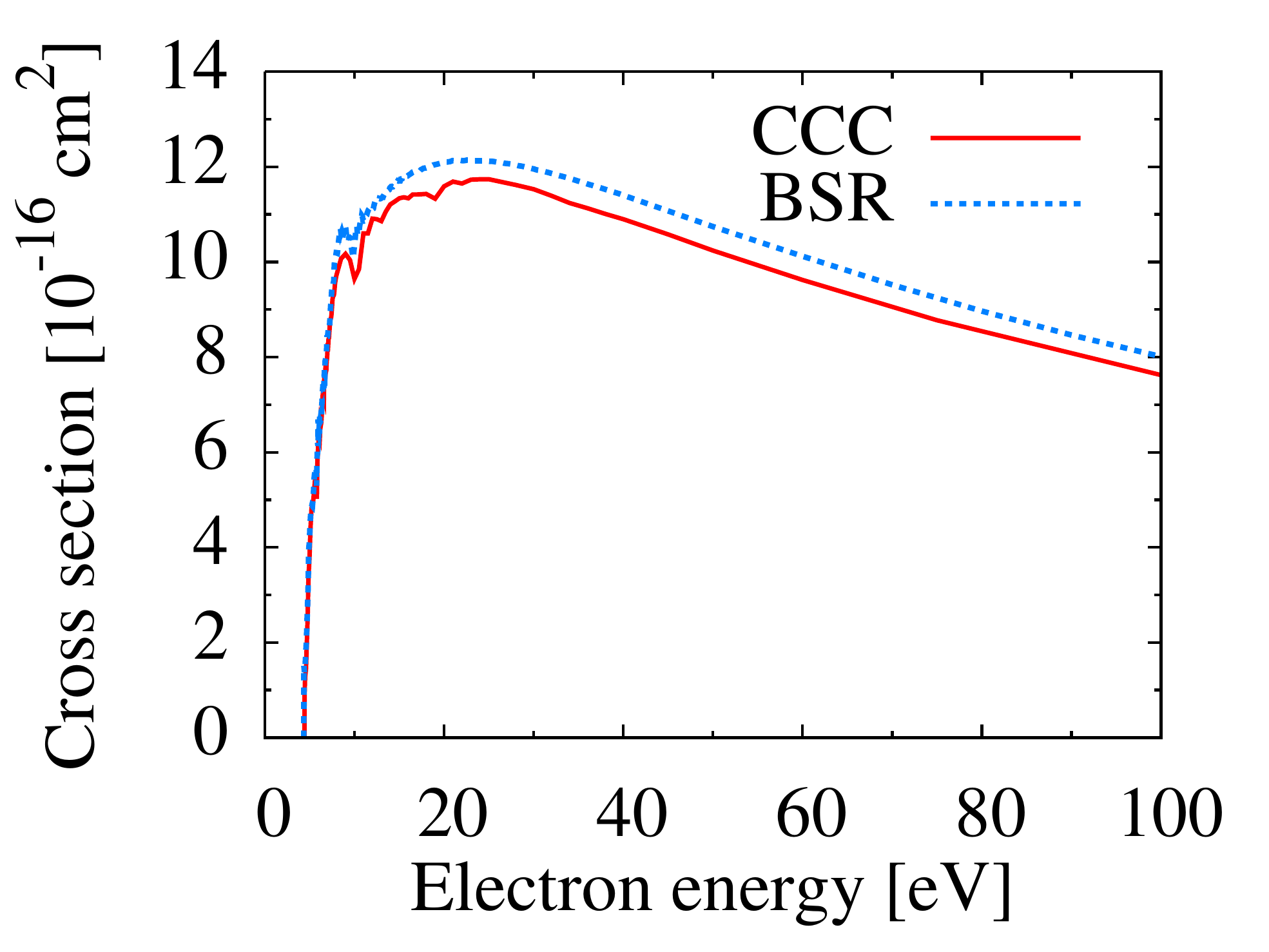}\put(40,20){\tiny 3s$^2$ $^1$S -- 3s3p $^1$P}\end{overpic}
   \begin{overpic}[width=0.24\textwidth]{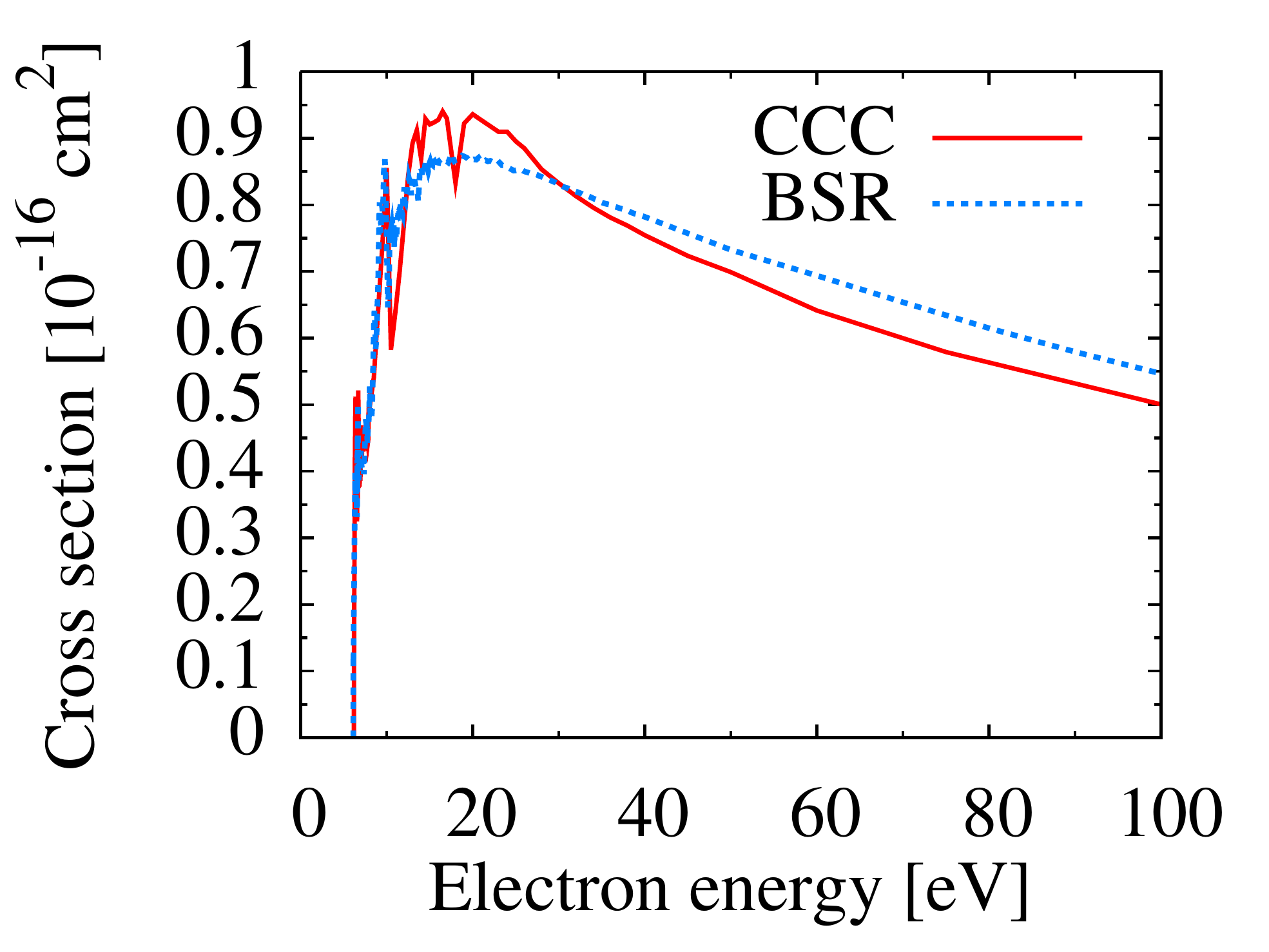}\put(40,20){\tiny 3s$^2$ $^1$S -- 3s4p $^1$P}\end{overpic}
   \begin{overpic}[width=0.24\textwidth]{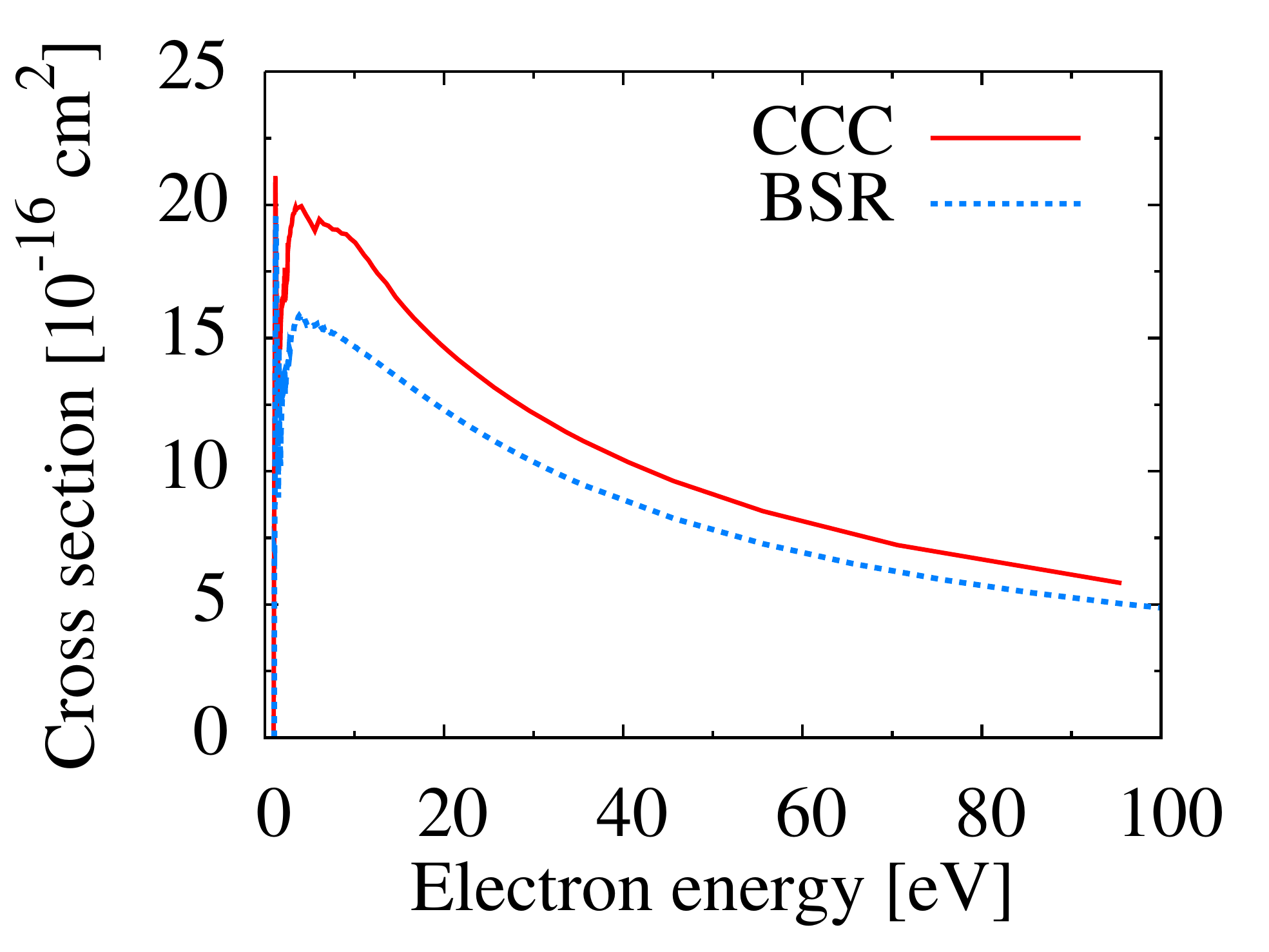}\put(40,20){\tiny 3s3p $^1$P -- 3s4s $^1$S}\end{overpic}
   \begin{overpic}[width=0.24\textwidth]{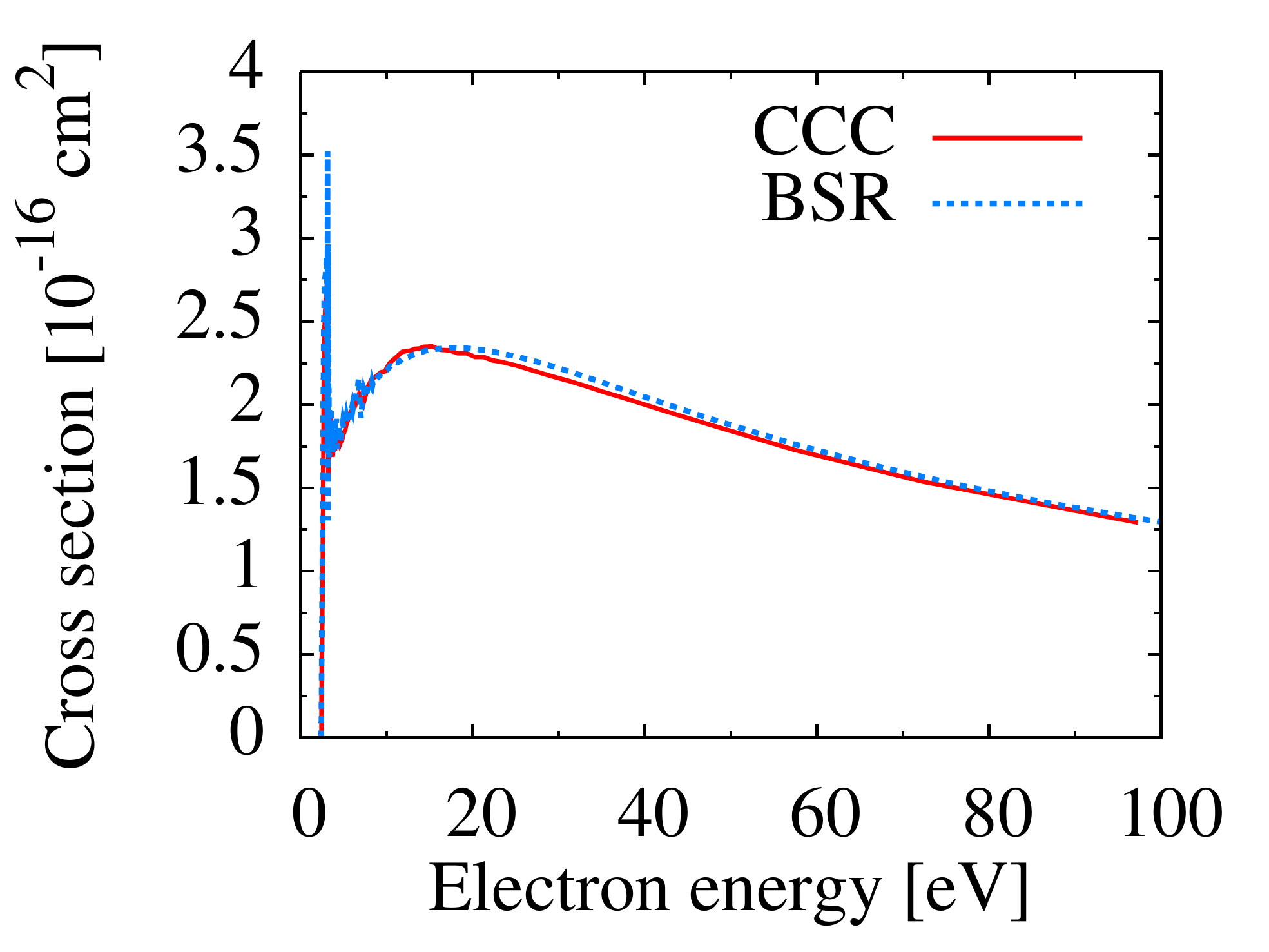}\put(40,20){\tiny 3s3p $^3$P -- 3s4s $^3$S}\end{overpic}
      \caption{Cross sections for selected electric dipole transitions.}
         \label{fig:cross_dipole}
   \end{figure}

\begin{figure}
   \centering
   \begin{overpic}[width=0.24\textwidth]{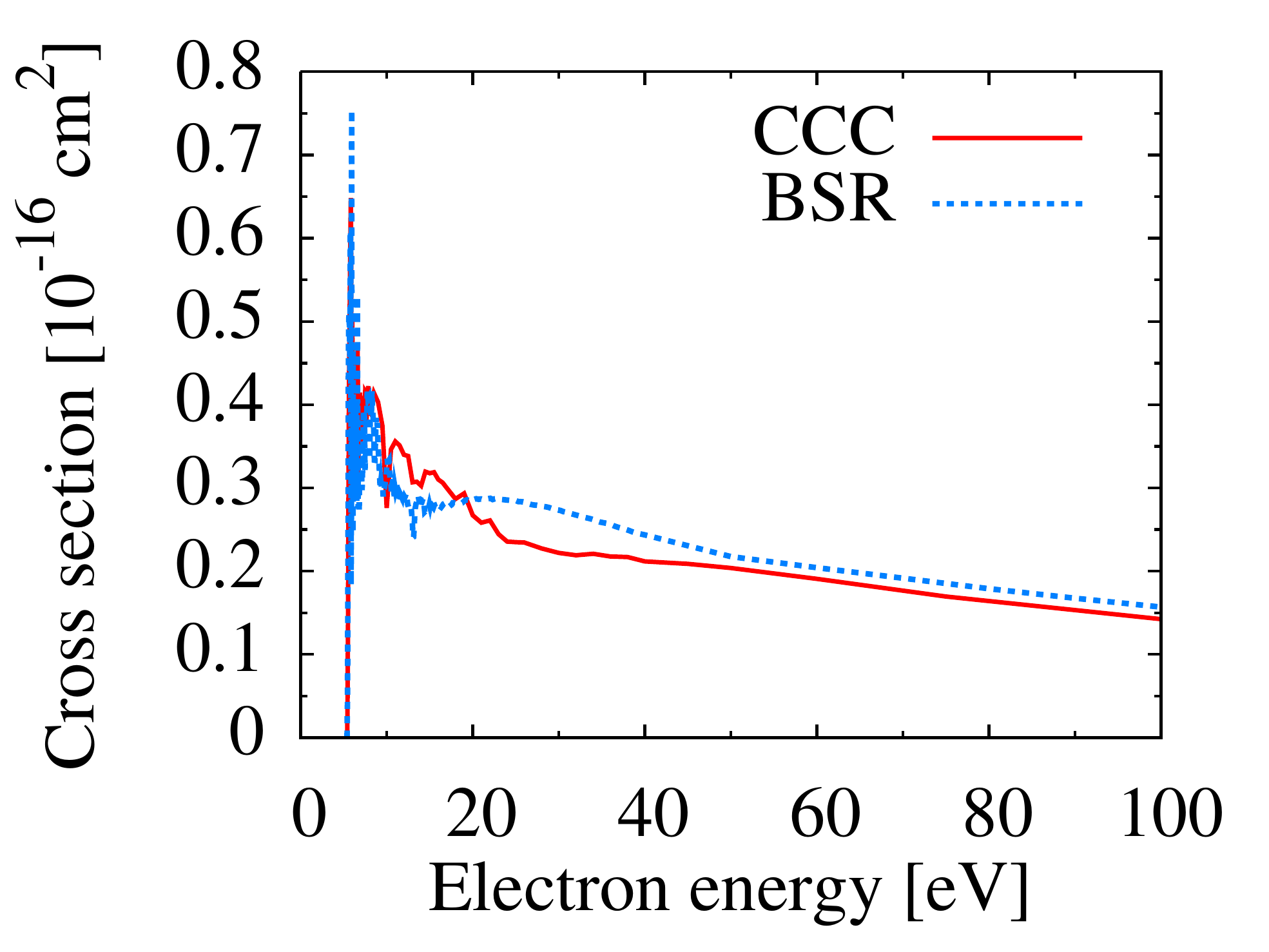}\put(30,20){\tiny 3s$^2$ $^1$S -- 3s4s $^1$S}\end{overpic}
   \begin{overpic}[width=0.24\textwidth]{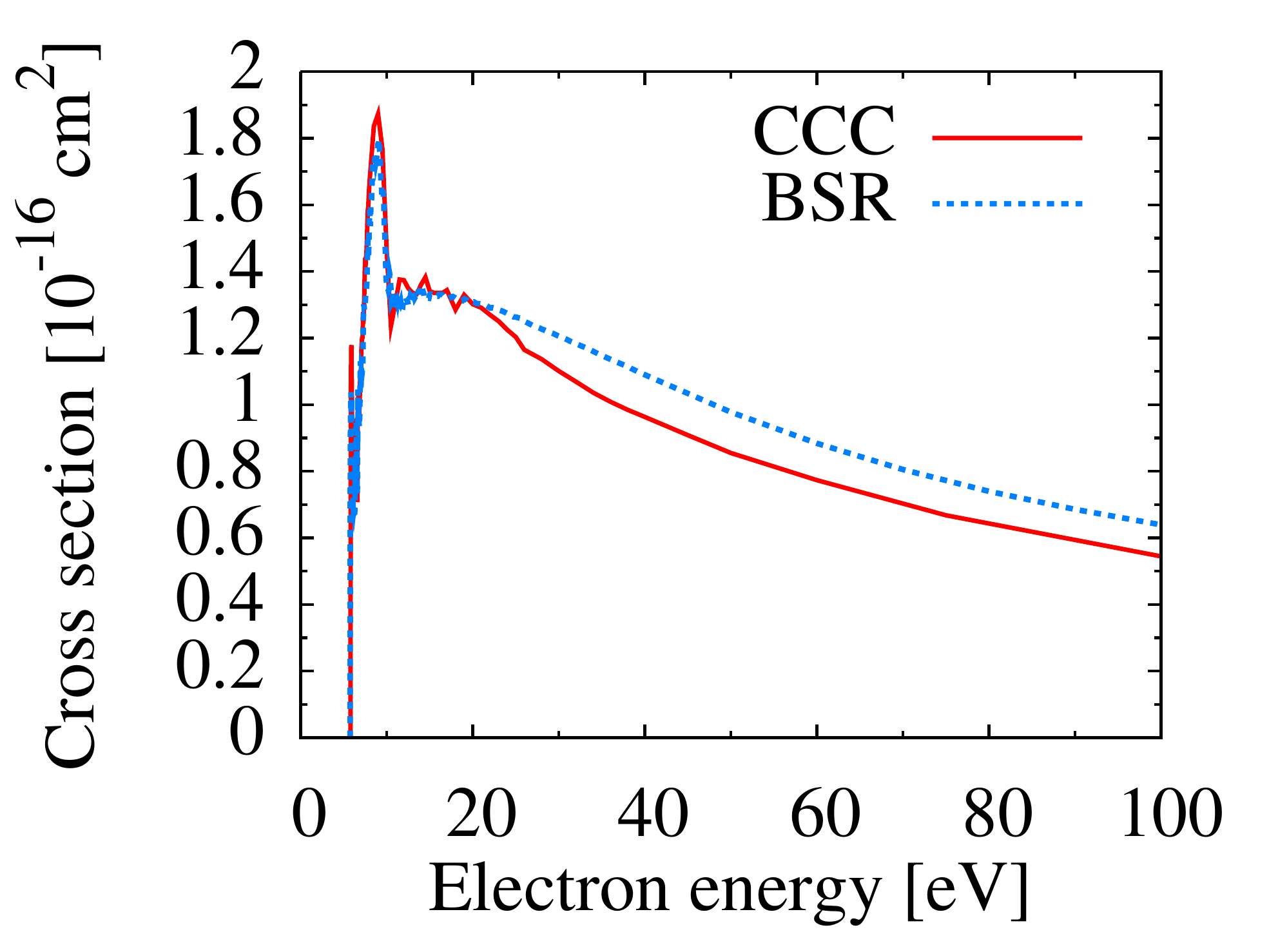}\put(30,20){\tiny 3s$^2$ $^1$S -- 3s3d $^1$D}\end{overpic}
   \begin{overpic}[width=0.24\textwidth]{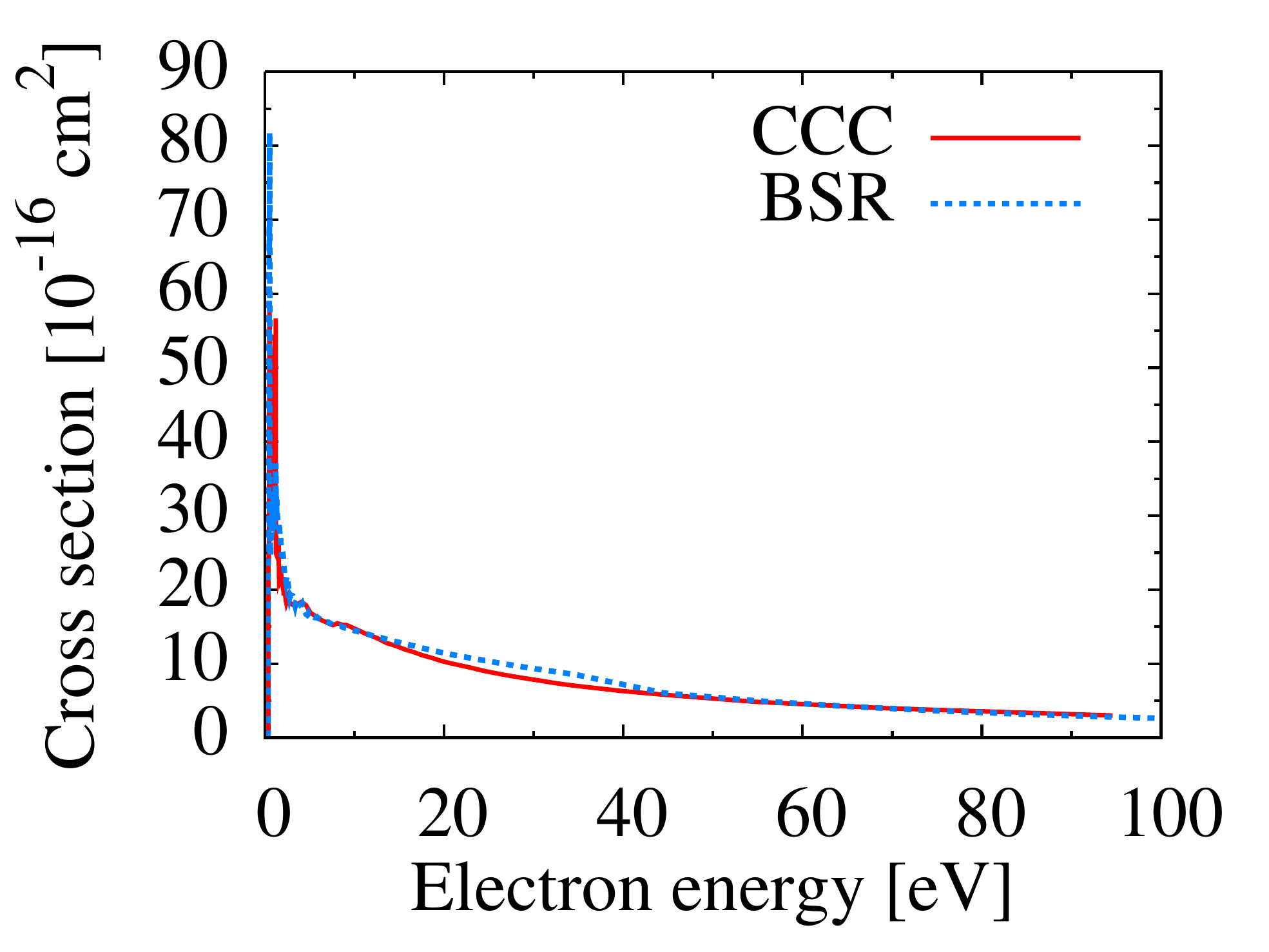}\put(30,35){\tiny 3s4s $^1$S -- 3s3d $^1$D}\end{overpic}
   \begin{overpic}[width=0.24\textwidth]{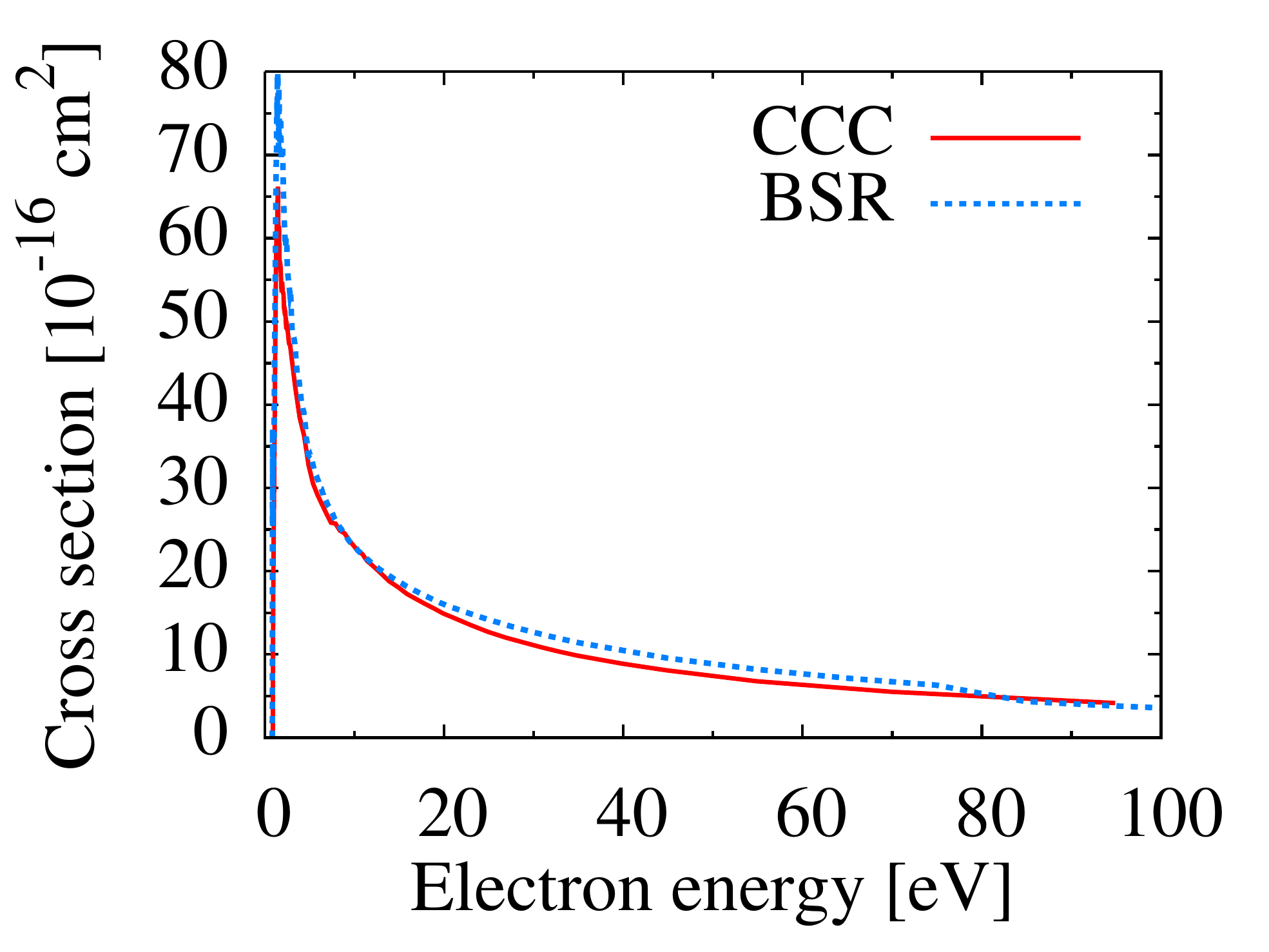}\put(30,35){\tiny 3s4s $^3$S -- 3s3d $^3$D}\end{overpic}
      \caption{Cross sections for selected spin-conserving non-electric dipole transitions. }
         \label{fig:cross_nondipole}
   \end{figure}

\begin{figure}
   \centering
   \begin{overpic}[width=0.24\textwidth]{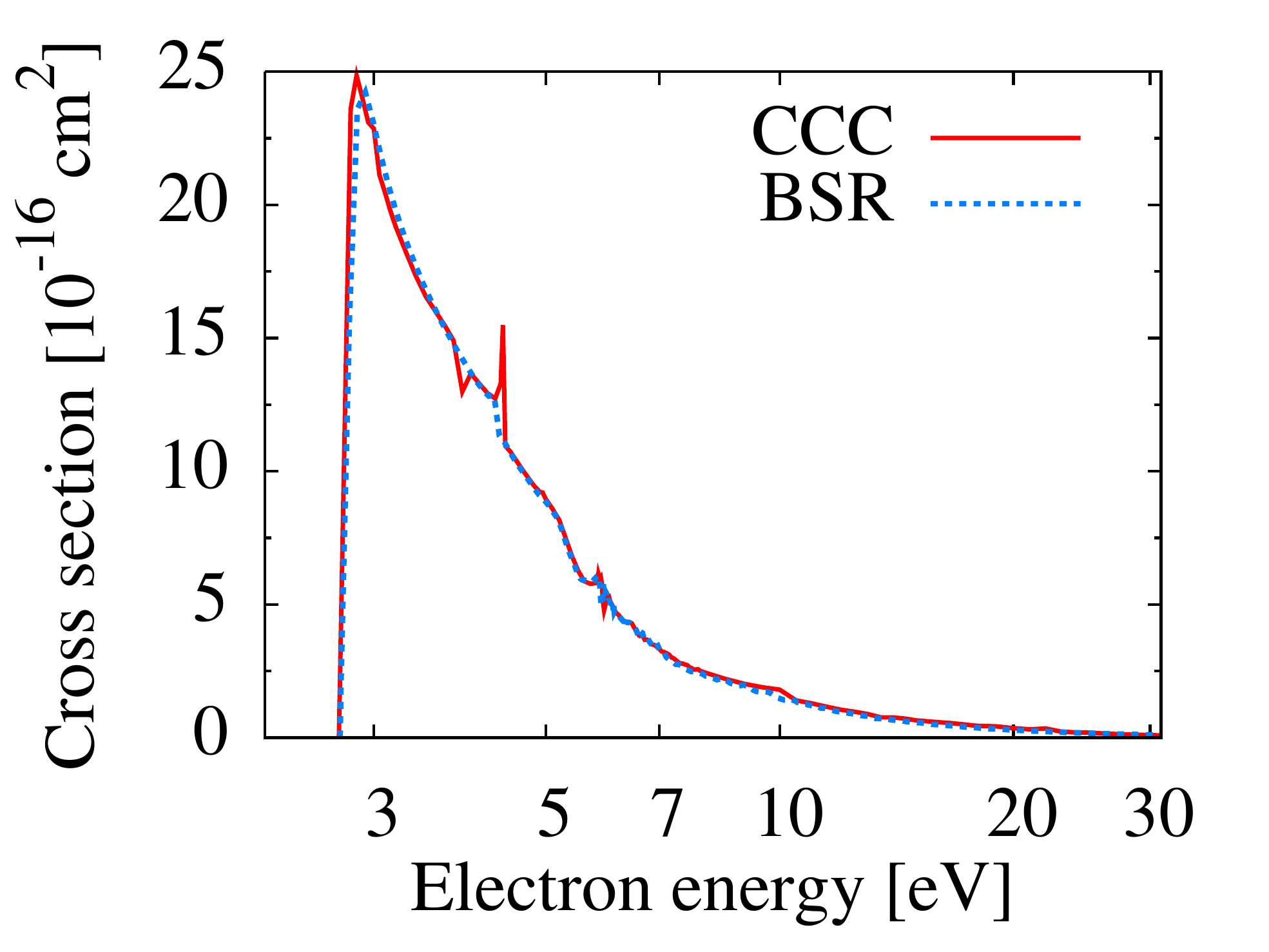}\put(45,40){\tiny 3s$^2$ $^1$S -- 3s3p $^3$P}\end{overpic}
   \begin{overpic}[width=0.24\textwidth]{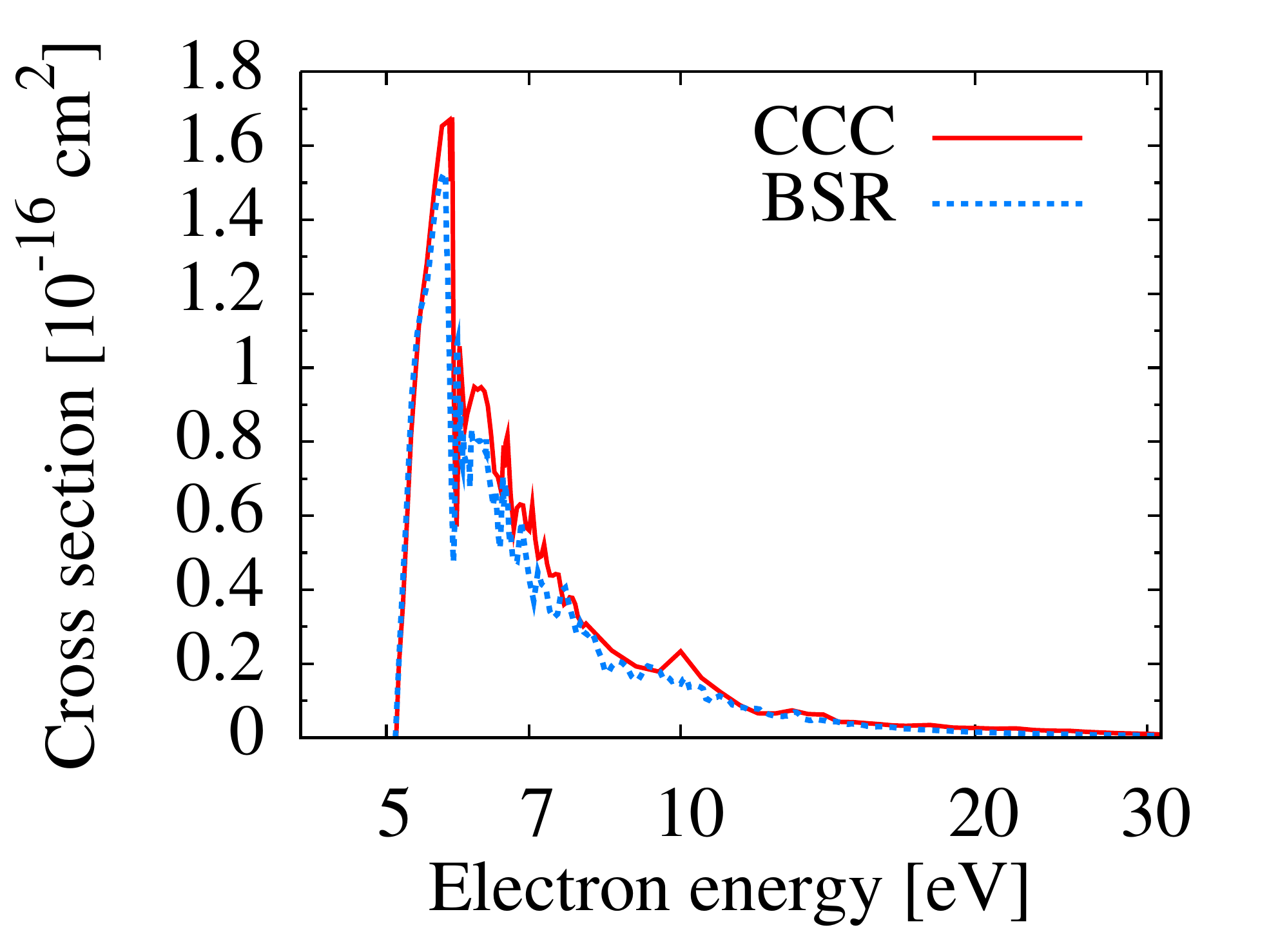}\put(45,40){\tiny 3s$^2$ $^1$S -- 3s4s $^3$S}\end{overpic}
   \begin{overpic}[width=0.24\textwidth]{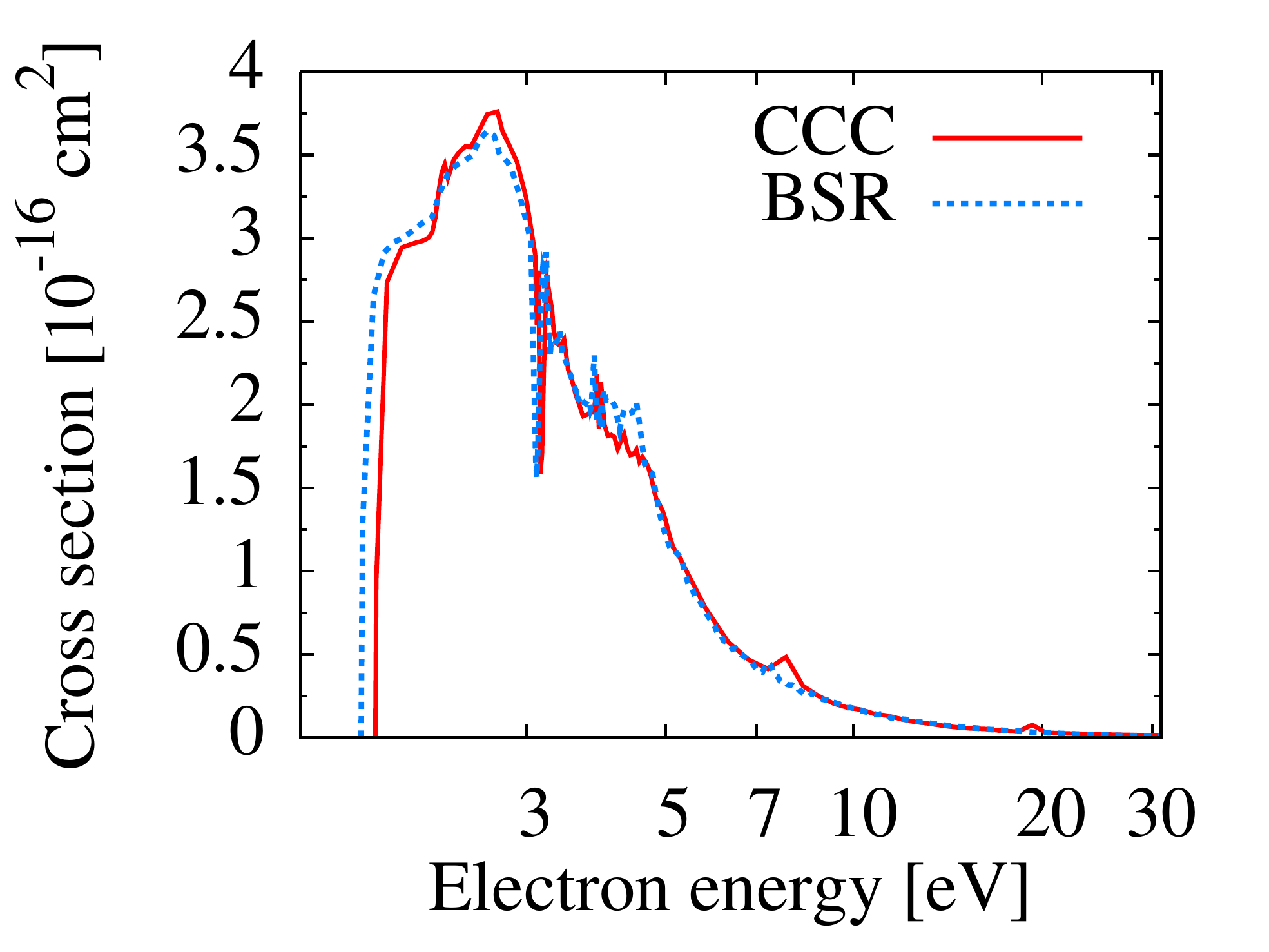}\put(60,40){\tiny 3s3p $^3$P}\put(60,30){\tiny -- 3s3p $^1$P}\end{overpic}
   \begin{overpic}[width=0.24\textwidth]{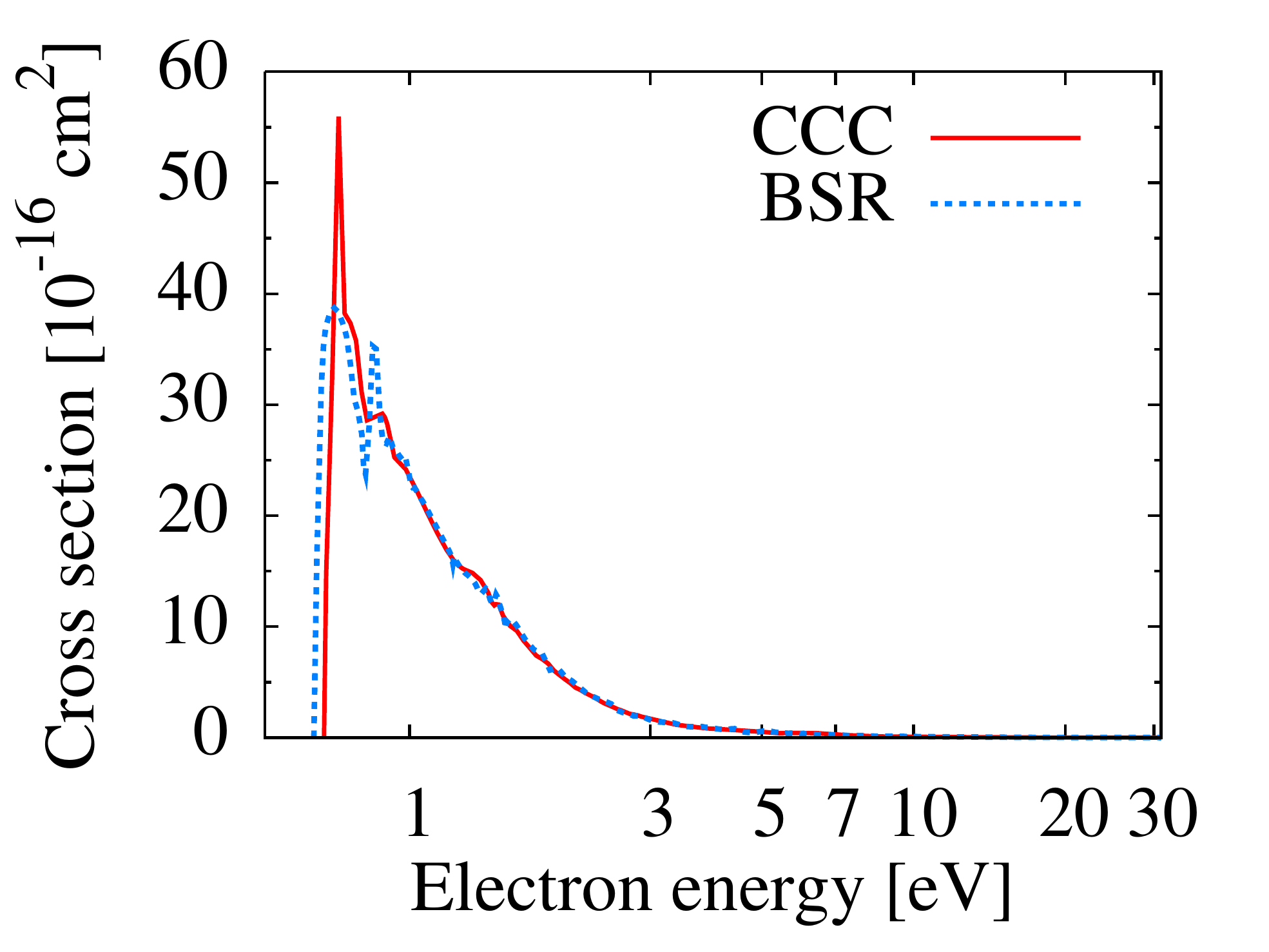}\put(40,40){\tiny 3s4s $^3$S -- 3s3d $^1$D}\end{overpic}
      \caption{Cross sections for selected spin-changing transitions.}
         \label{fig:cross_exchange}
   \end{figure}

In both the CCC and BSR calculations, the 25 lowest contiguous physical states of Mg, 
up to 3s6p$\ ^1$P, fit into the effective boxes defined in the computational models (either by
the exponent in the Laguerre basis or by the R-matrix radius), and robust cross sections obtained.  
%%The individual 
%%calculations include a few higher states , separated from the lowest 25 (i.e.\ some 
%%intermediate physical states are not modelled); however, to be able to make a 
%%coherent comparison and evaluation of the reliability of the results, 
%The present discussion is limited to these lowest 25 states.  
Table~\ref{tab:energies} 
presents the considered states, along with their experimental excitation energies.  
In both cases, the collision calculations were performed for electron energies from 
threshold to about 100~eV.  Cross sections for selected cases of electric dipole, 
non-dipole spin-conserving, and spin-changing transitions are shown in 
Figs.~\ref{fig:cross_dipole},~\ref{fig:cross_nondipole}, and~\ref{fig:cross_exchange}, 
respectively.  Note, BSR structure calculations for Mg  were presented and discussed in \cite{zatsarinny_cross_2009}.
Overall, they present an accurate description of the binding energies and oscillator strengths.  However, in the scattering calculations the experimental excitation energies are adjusted to the experimental ones in Table~\ref{tab:energies} to remove possible additional uncertainties related to the error in the excitation thresholds.
In the CCC calculations, the energies from the structure calculations are retained and therefore the cross section comparisons in the figures checks both the target structure and collision dynamics of the calculations.

\begin{table}
\begin{center}
\caption{Physical states for which data are provided from the CCC and BSR calculations, 
along with experimental excitation energies ($E_\mathrm{expt}$) from NIST \citep{NIST_5.4, martin_energy_1980}.}
\label{tab:energies}
\begin{tabular}{rrrr}
\hline \hline
Index & State & g & $E_\mathrm{expt}$ \\
      &       &   &  [eV]             \\
\hline
  1 &  3s$^2$ $^1$S &   1 &   0.000  \\
  2 &    3s3p $^3$P &   9 &   2.714  \\
  3 &    3s3p $^1$P &   3 &   4.346  \\
  4 &    3s4s $^3$S &   3 &   5.108  \\
  5 &    3s4s $^1$S &   1 &   5.394  \\
  6 &    3s3d $^1$D &   5 &   5.753  \\
  7 &    3s4p $^3$P &   9 &   5.932  \\
  8 &    3s3d $^3$D &  15 &   5.946  \\
  9 &    3s4p $^1$P &   3 &   6.118  \\
 10 &    3s5s $^3$S &   3 &   6.431  \\
 11 &    3s5s $^1$S &   1 &   6.516  \\
 12 &    3s4d $^1$D &   5 &   6.588  \\
 13 &    3s4d $^3$D &  15 &   6.719  \\
 14 &    3s5p $^3$P &   9 &   6.726  \\
 15 &    3s4f $^1$F &   7 &   6.779  \\
 16 &    3s4f $^3$F &  21 &   6.779  \\
 17 &    3s5p $^1$P &   3 &   6.783  \\
 18 &    3s6s $^3$S &   3 &   6.930  \\
 19 &    3s6s $^1$S &   1 &   6.966  \\
 20 &    3s5d $^1$D &   5 &   6.981  \\
 21 &    3s5d $^3$D &  15 &   7.063  \\
 22 &    3s6p $^3$P &   9 &   7.069  \\
 23 &    3s5f $^1$F &   7 &   7.092  \\
 24 &    3s5f $^3$F &  21 &   7.092  \\
 25 &    3s6p $^1$P &   3 &   7.094  \\
 \hline 
\end{tabular}
\end{center}
\end{table} 

The present CCC and BSR calculations are performed in a nonrelativistic formulation of these methods.  \ion{Mg}{I} is a light atom and it is expected that relativistic effects  are small for both collision dynamics and target wave functions.  As a consequence, excitation of \ion{Mg}{I} triplet states proceeds via exchange scattering.  We have verified this by performing fully relativistic CCC calculations \citep{fursa_fully_2008} at a number of energies and finding good agreement with nonrelativistic CCC and BSR results.  A comparison of relativistic and nonrelativistic cross sections requires a transformation from $LS$ to $LSJ$ coupling for the nonrelativistic theory and introduction of statistical weights (1/3 for the $3s3p^3$P state) that multiply the nonrelativistic cross sections.
A simple way to estimate the importance of relativistic effects is to evaluate how they affect the target wave functions. For example, for the $3s3p^3$P$_1$ state this can be approximately described by the singlet-triplet mixing of the nonrelativistic wave functions with a mixing coefficient of order $10^{-3}$.  The effect on the cross sections is proportional to the square of the mixing coefficient and is negligible at all incident electron energies considered in this paper.
Note, however, that the radiative intercombination $3s^2 \, ^1$S - $3s3p \, ^3$P$_1$ line is a result of relativistic effects.
   
In applications, the rate coefficient is required, which is calculated by folding the 
cross sections $\sigma$ with the velocity distribution, assumed here to be the Maxwell 
distribution.  That is
\begin{equation}
\langle \sigma \varv \rangle = \left(\frac{8}{\pi\mu}\right)^{1/2} \frac{1}{(kT)^{3/2}} \int_{\Delta E}^\infty \sigma(E) E \exp\left( - \frac{E}{kT}\right) dE,
\label{eq:ratecoeff}
\end{equation}
where $\mu$ is the reduced mass of the colliding system, $\Delta E$ is the energy 
threshold for the given process, $E$ is the collision energy (in the centre-of-mass frame), 
and other symbols have their usual meanings.  Following \cite{Seaton1953}, one can define 
the thermally-averaged or \emph{effective} collision strength, which is related to the 
rate coefficients by simple expressions.  In cgs units, for $E_j>E_i$, and thus for 
excitation we have
\begin{equation}
\langle \sigma_{ij}\varv \rangle= 8.63 \times 10^{-6} \frac{\Upsilon_{ij}(T)}{g_i \sqrt{T}} e^{-\Delta E/kT},
\label{eq:collstr}
\end{equation}
where $\Delta E = E_j-E_i$, and for deexcitation (via the detailed balance relation),
\begin{equation}
\langle \sigma_{ji}\varv \rangle= 8.63 \times 10^{-6} \frac{\Upsilon_{ij}(T)}{g_j \sqrt{T}}.
\end{equation}
The collision strengths have the advantage of being both dimensionless and symmetric.  
We have therefore chosen to present and discuss effective collision strengths $\Upsilon_{ij}$, 
even though rate coefficients $\langle \sigma_{ij} \varv \rangle$ are required 
in non-LTE applications.  The calculation of the effective collision strengths was 
done in all cases by first calculating the rate coefficient from the excitation 
cross sections via Eq.~(\ref{eq:ratecoeff}), and then transforming to the effective 
collision strength using Eq.~(\ref{eq:collstr}), using the data in Table~\ref{tab:energies}.  

%-----------------------------------------------------------------
\section{Results and discussion}

The results of the calculations for the effective collision strengths $\Upsilon_{ij}$ 
from the CCC and BSR calculations are presented for temperatures $T$ ranging from 
1000~K to 10000~K in steps of 1000~K, with additional results for 500~K at the cool end, 
and 15000~K and 20000~K at the hot end.  This temperature grid is chosen to focus on the 
range of temperatures where the abundance of neutral magnesium will be significant.  
At cooler temperatures, there will be significant molecule formation, and at hotter 
temperatures Mg will be ionised.  The data are presented as matrices, following the 
ordering of indexes in Table~\ref{tab:energies} (i.e. the transition 1--2 corresponds 
to element (1,2)), one for each temperature.  An example matrix each for the CCC and 
BSR calculations at 5000~K is shown in tables~\ref{tab:upsilon_ccc} and~\ref{tab:upsilon_bsr}, 
respectively.  The full tables, and tables for other temperatures, are available electronically at CDS.

\begin{table*}
\begin{center}
\caption{Sample of electronic table for effective collision strengths $\Upsilon_{ij}$ 
at 5000~K from the CCC calculation.  The full table, and tables for other temperatures, 
are available electronically at CDS.  }
\label{tab:upsilon_ccc}
\begin{verbatim}   
 0.00e+00  4.84e+00  9.74e-01  4.08e-01  1.65e-01  4.39e-01  2.47e-01   ...   
 4.84e+00  0.00e+00  3.87e+00  3.46e+00  6.50e-01  3.10e+00  4.52e+00   ...   
 9.74e-01  3.87e+00  0.00e+00  2.38e+00  6.11e+00  8.29e+00  3.92e+00   ...   
 4.08e-01  3.46e+00  2.38e+00  0.00e+00  6.05e-01  5.11e+00  1.49e+01   ...   
 1.65e-01  6.50e-01  6.11e+00  6.05e-01  0.00e+00  2.31e+00  1.64e+00   ...   
 4.39e-01  3.10e+00  8.29e+00  5.11e+00  2.31e+00  0.00e+00  1.22e+01   ...   
 2.47e-01  4.52e+00  3.92e+00  1.49e+01  1.64e+00  1.22e+01  0.00e+00   ...   
  ...       ...       ...       ...       ...       ...       ...       ...         
\end{verbatim}
\end{center}
\end{table*} 

\begin{table*}
\begin{center}
\caption{Sample of electronic table for effective collision strengths $\Upsilon_{ij}$ 
at 5000~K from the BSR calculation.  The full table, and tables for other temperatures, 
are available electronically at CDS.  }
\label{tab:upsilon_bsr}
\begin{verbatim}   
 0.00e+00  4.65e+00  1.16e+00  3.87e-01  1.71e-01  4.32e-01  2.63e-01   ...   
 4.65e+00  0.00e+00  4.52e+00  3.85e+00  7.04e-01  3.14e+00  5.29e+00   ...   
 1.16e+00  4.52e+00  0.00e+00  2.08e+00  4.77e+00  7.68e+00  3.41e+00   ...   
 3.87e-01  3.85e+00  2.08e+00  0.00e+00  6.27e-01  5.36e+00  1.51e+01   ...   
 1.71e-01  7.04e-01  4.77e+00  6.27e-01  0.00e+00  2.15e+00  1.93e+00   ...   
 4.32e-01  3.14e+00  7.68e+00  5.36e+00  2.15e+00  0.00e+00  1.07e+01   ...   
 2.63e-01  5.29e+00  3.41e+00  1.51e+01  1.93e+00  1.07e+01  0.00e+00   ...     
  ...       ...       ...       ...       ...       ...       ...       ...         
\end{verbatim}
\end{center}
\end{table*} 

\begin{figure}
   \centering
   \includegraphics[width=90mm,angle=0]{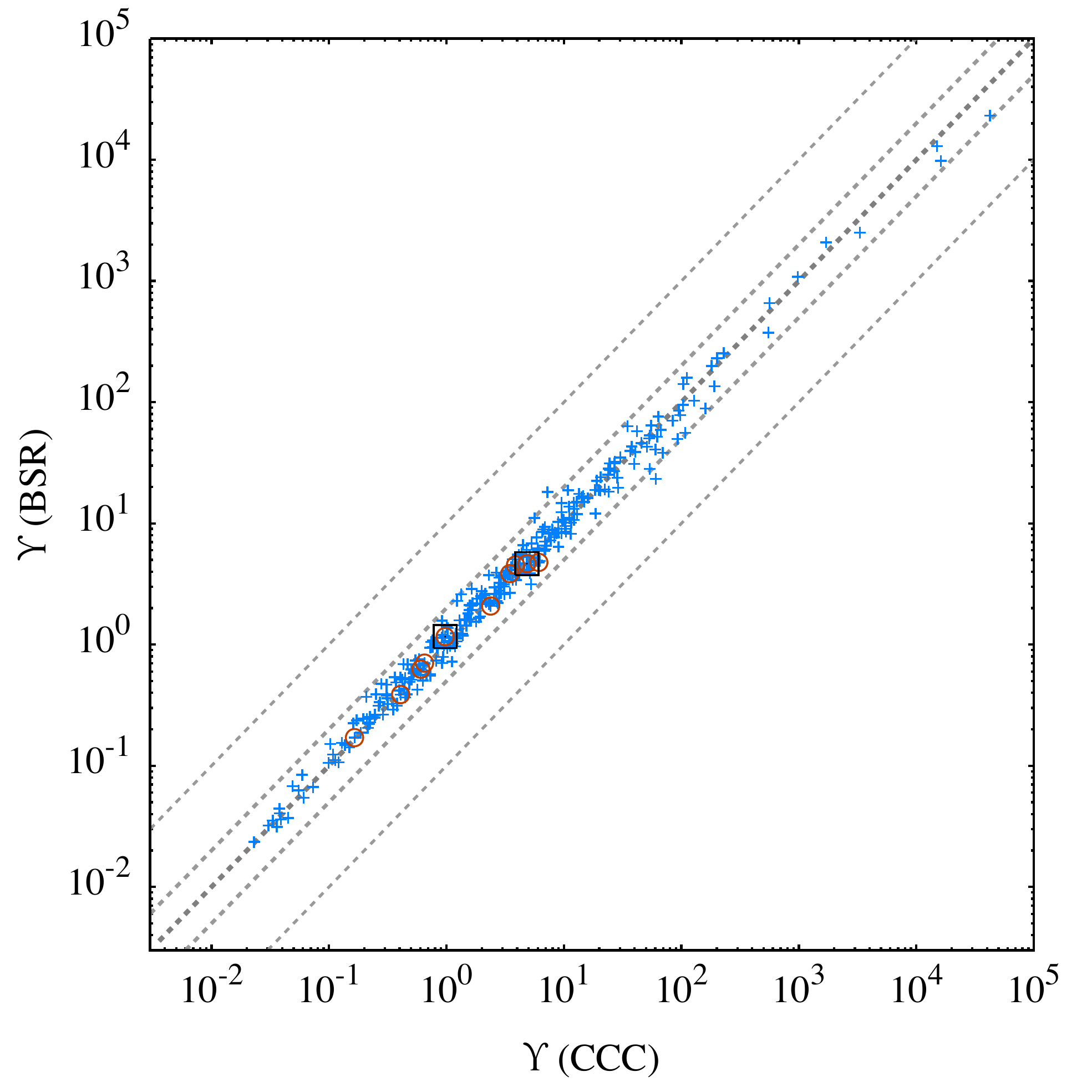}
      \caption{Comparison of the effective collision strengths $\Upsilon_{ij}$ at 
      5000~K from CCC and BSR.  The central dashed line shows the one-to-one relation, with 
      the outer dashed lines showing factors of 2 and 10 departures from this relation.  
      Circled points are transitions involving the lowest 5 states of Mg.  
      Points shown surrounded by a square are the transitions from the ground 
      state to the two lowest-lying excited states, i.e.\  
      3s$^2 \, ^1\mathrm{S}$ -- 3s3p$ \, ^3\mathrm{P}$ and  
      3s$^2 \, ^1\mathrm{S}$ -- 3s3p$ \, ^1\mathrm{P}$, 
      the former always having a larger effective collision strength.   }
         \label{fig:comp}
   \end{figure}

The results from the CCC and BSR calculations at 5000~K are compared in Fig.~\ref{fig:comp}.  
The agreement between the calculations is quantified in Table~\ref{tab:stats}, 
where statistical measures of the location (offset) and scale (scatter) of the ratios 
of thermally averaged collision strengths are given.   The agreement between the CCC 
and BSR calculations is very good, with an offset of only 5\% (BSR larger than CCC), 
and a scatter of less than 30\%. The agreement between the CCC and BSR results, from 
independent methods and calculations, suggests that these datasets are reliable to 
this level of accuracy. 

\begin{table}
\begin{center}
\caption{The location and scale of the ratio $\Upsilon_{ij}(1)/\Upsilon_{ij}(2)$, 
where $\Upsilon_{ij}(1)$ corresponds to dataset $1$, and $\Upsilon_{ij}(2)$ 
corresponds to dataset $2$.  The ratio is assumed to follow a log-normal distribution 
and thus the location (offset) and scale (scatter) are characterised by the geometric 
mean $e^\mu$ and the geometric coefficient of variation $e^\sigma-1$, where $\mu$ and 
$\sigma$ are the mean and standard deviations of $\log(\Upsilon_{ij}(1)/\Upsilon_{ij}(2)$), 
respectively. These measures are used since the comparisons with \cite{Mauas1988} show 
skewness, implying a log-normal distribution. }
\label{tab:stats}
\begin{tabular}{lrrr}
\hline \hline
$_{\mathrm{dataset}\,2} \backslash ^{\mathrm{dataset}\,1}$ & CCC  & BSR & M88 \\
\hline
 & \multicolumn{3}{c}{\underline{location}} \\
CCC                    &      1.00   &   1.04   &   0.44  \\
BSR                    &      0.96   &   1.00   &   0.42  \\
M88                    &      2.26   &   2.40   &   1.00  \\
 & \multicolumn{3}{c}{\underline{scale}} \\
CCC                    &      0.00   &   0.27   &   6.57  \\
BSR                    &      0.27   &   0.00   &   6.40  \\
M88                    &      6.57   &   6.40   &   0.00  \\
\hline 
\end{tabular}
\end{center}
\end{table} 
       
\begin{figure}
   \centering
   \includegraphics[width=90mm,angle=0]{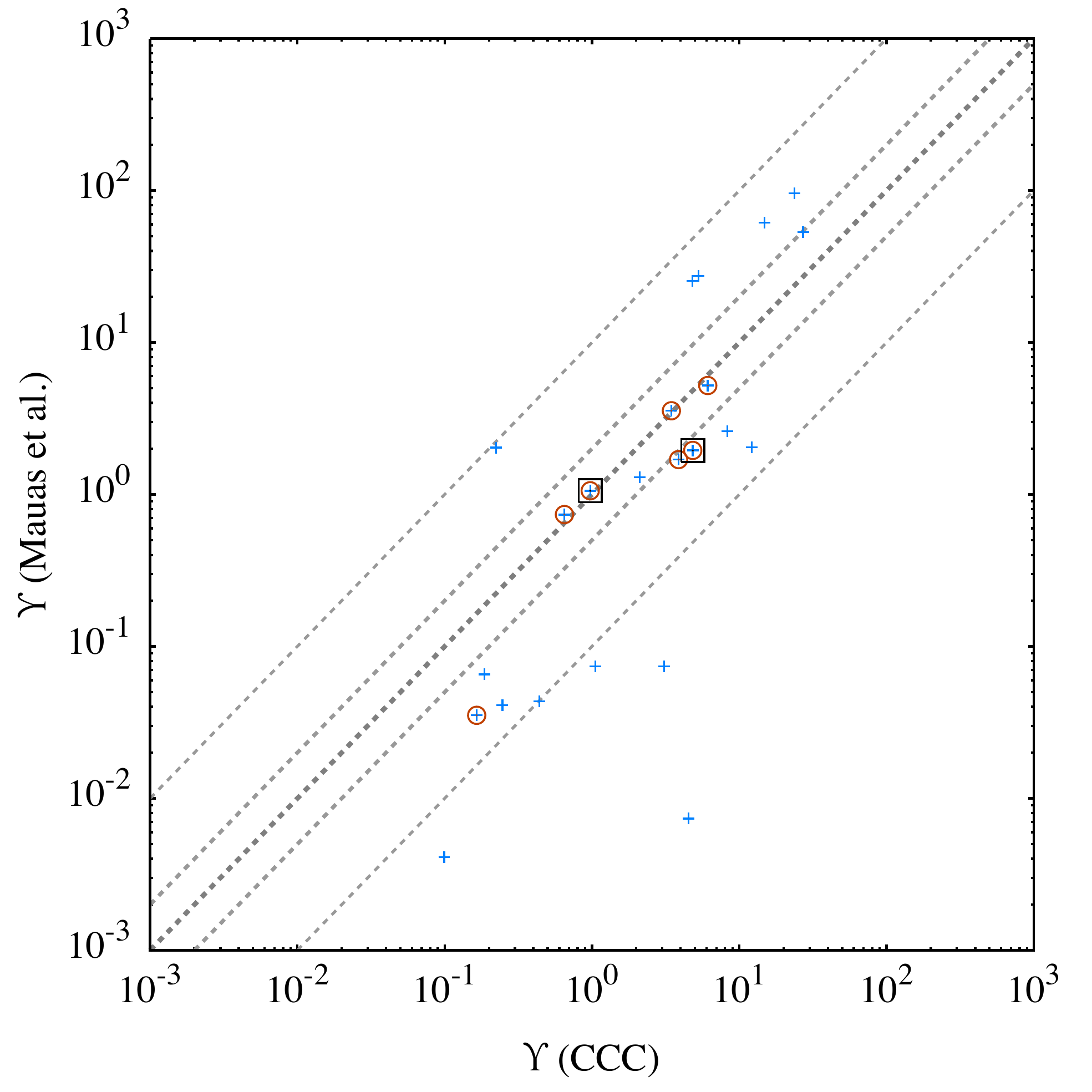}
      \caption{Comparison of the effective collision strengths $\Upsilon_{ij}$ 
      from \cite{Mauas1988}, with those from the CCC calculations, at 5000~K.  
      The lines and points follow the description in Fig.~\ref{fig:comp}. }
         \label{fig:ccc_mauas}
   \end{figure}

As mentioned in \S~1, the set of electron collision data assembled by \cite{Mauas1988} (M88)
is widely used in astro\-physical applications.  It combines data from theory, usually 
in the Born approximation, experimental results where available, and results from the 
semi-empirical Bethe-Born formula of \cite{vanRegemorter1962} for optically allowed 
transitions, and an arbitrary scaling of the \citeauthor{vanRegemorter1962} formula 
for forbidden transitions.  Figure~\ref{fig:ccc_mauas} shows the comparison of the 
effective collision strengths corresponding to the data from M88 with the CCC results at 5000~K, 
and the statistical comparisons with other datasets also quantified in \hbox{Table~\ref{tab:stats}}.  
It is seen that the agreement is usually within a factor of 10, but with some cases 
disagreeing by significantly more.\footnote{This comparison only covers the data tabulated in Tables 2 and 3 of \cite{Mauas1988}.  The arbitrary values listed in the note to Table 3 are not included.  These would make the comparison significantly worse.} The transition between the two lowest 
states, 3s$^2 \, ^1\mathrm{S}$ -- 3s3p$ \, ^3\mathrm{P}$, is of particular 
relevance, as it corresponds to the intercombination line at 457~nm, and thus 
controls this transition.  We note that the value in M88 is 
significantly lower, only 40\% of the values found in the 
CCC and BSR calculations i.e.\ a factor of 2.5 lower.  

The precise origin of this discrepancy is difficult to determine.  \citeauthor{Mauas1988} state that to obtain the data for this transition they ``combined the absolute value of \cite{robb_born_1974} and the energy dependence measured by \cite{aleksakhin_electron-impact_1973}.''  The mentioned paper by \citeauthor{robb_born_1974} presents Born cross sections for the resonance 3s$^2 \, ^1\mathrm{S}$ -- 3s3p$ \, ^1\mathrm{P}$ transition at electron energies above 10 eV, but does not cover the 3s$^2 \, ^1\mathrm{S}$ -- 3s3p$ \, ^3\mathrm{P}$ transition.   We note that another paper by the same author, \cite{robb_calculations_1975}, employs the $R$-matrix method and does indeed include results for the 3s$^2 \, ^1\mathrm{S}$ -- 3s3p$ \, ^3\mathrm{P}$ transition; these results are reproduced by \citet[][see Fig.~21]{fabrikant_electron_1988}.   This calculation obtains a maximum cross section of about $22 \times 10^{-16}$~cm$^2$, while the present results have a maximum cross section of about $25 \times 10^{-16}$~cm$^2$; see Fig.~\ref{fig:cross_exchange}.  However, we note that the energy dependence measured by \cite{aleksakhin_electron-impact_1973} obtains its maximum value at $\sim 4.5$~eV, with significantly lower values near the threshold at 2.7~eV.  In contrast, the present results find the maximum value very close to threshold, 2.7--3.0~eV, and the cross section has already decreased to half the maximum value by $\sim 4.5$~eV.   Thus, it seems likely that the discrepancy primarily originates in the different threshold behaviours of the present calculations compared to those measured and presented by \cite{aleksakhin_electron-impact_1973}.  It is unclear to what degree the details of these experimental results are reliable.  We also note that the threshold behaviour for the 3s$^2 \, ^1\mathrm{S}$ -- 3s3p$ \, ^1\mathrm{P}$ transition obtained by \citeauthor{aleksakhin_electron-impact_1973} is also considerably less steep than that measured by \cite{leep_excitation_1976} (see Fig.~2 of that paper), which may indicate an appreciable uncertainty.
   
Comparisons with the data of \cite{merle_effective_2015} and \cite{osorio_mg_2015} were also performed, but for the sake of simplicity of presentation, we have chosen not to present or analyse these in detail.  It suffices to say that these datasets compare significantly better than M88 with the new data, with agreement to better than a factor of two.  However, the agreement is certainly worse than the internal agreement of the present CCC and BSR results.%, and thus again we may conclude that the new datasets are to be preferred.

%-----------------------------------------------------------------
\section{Applications}

To make an initial assessment of the impact of the new data, we performed a small number of calculations in example astrophysical applications: in late-type stellar atmosphere models, and in a supernova ejecta model.  %These examples show important effects on the 457~nm intercombination line.

\subsection{Late-type stellar atmospheres}

To test the effects on modelling of Mg lines in late-type stellar atmospheres, we computed synthetic non-LTE spectra following \cite{osorio_mg_2015}.  As in that paper, we employ four atmospheric models from the MARCS grid \citep{Gustafsson2008}: two giants  (\teff (K), \logg = 4500, 1.5) and two dwarfs (\teff(K), \logg = 6000, 4.5); in each case one with solar metallicity, \feh= 0.0, and the other metal-poor, \feh = $-2.0$.  The model atoms used are identical to the final model atom in \citeauthor{osorio_mg_2015} (model ``F'') differing only in the electron collision data.  For collisions between the states shown in Table~\ref{tab:energies} we used the CCC data in one version of the model atom, and BSR in another. For the remaining transitions, the impact parameter (IP) formula from \cite{Seaton1962b} was used. Additionally, we constructed a model atom with the electron collision data from M88 and again used the IP formula for the remaining transitions.  \citeauthor{osorio_mg_2015} should be consulted for additional details regarding the model atom and the non-LTE radiative transfer calculations.

\begin{figure}
\includegraphics[width=0.5\textwidth]{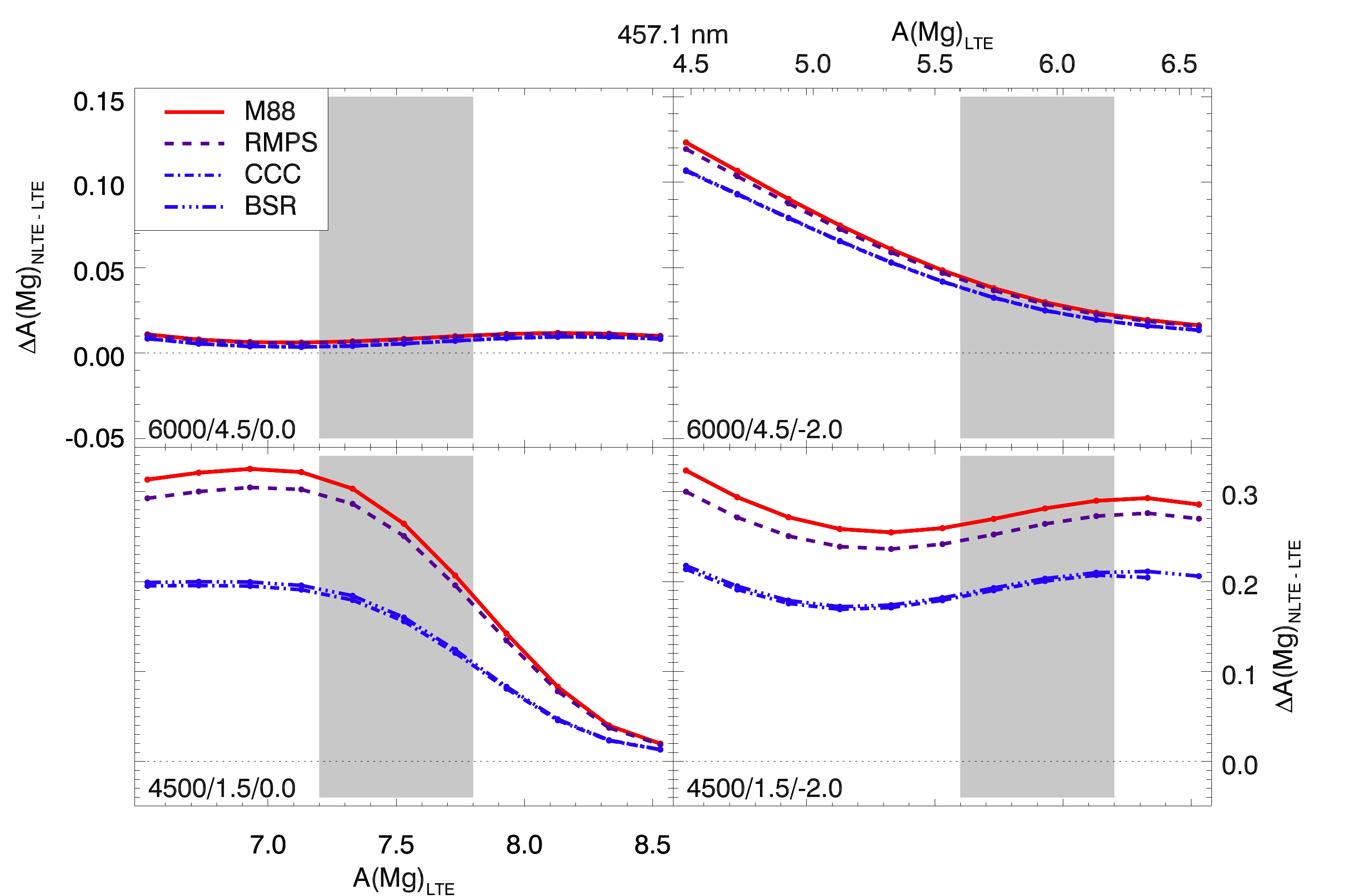}
\caption{Abundance corrections for the 457~nm line using four atmospheric models and four model atoms differing only in the electron collisional data used. The parameters of the atmospheric models are shown in the bottom left of each panel in the format \teff/\logg/\feh. The M88 model atom (solid line) includes data from \cite{Mauas1988}, the RMPS model is the model ``F'' used in \cite{osorio_mg_2015} (dashed line), the CCC model uses the CCC data presented in this work (dot-dashed line) and the BSR model uses the BSR data presented here (triple-dot-dashed line). The shaded region shows the expected values of A(Mg) for a typical star of the given metallicity. The black points show the calculated points.}\label{fig:acorr}
\end{figure}

All the spectral lines considered in Table~1 of \citeauthor{osorio_mg_2015} were computed, and the effects examined.  The new data have no effect in any of the models on the infrared lines at 7, 12 and 18~$\mu$m, as would be reasonably expected from the fact that these lines arise from levels lying higher than those for which new data is available.  Regarding the absorption lines at $\lambda < 1 \, \mu$m, we examined the effects on the final abundance corrections\footnote{Defined as $\Delta \rm{A(Mg)}= \rm{A(Mg)}_{\rm{_{non-LTE}}} - \rm{A(Mg)}_{\rm{_{LTE}}}$, with $\rm{A(Mg)}=\log(N_{_{Mg}}/N_{_{H}})+12 $} $\Delta \rm{A(Mg)}$, and for all lines these were found to be $<$ 0.01 dex in all four stellar models, except in the case of the 457~nm line.  The abundance corrections for this line for the four atmospheric models, using the various electron collision datasets, are shown in Fig.~\ref{fig:acorr}.  For both dwarf models, this line also shows negligible differences in $\Delta \rm{A(Mg)}$ regardless of the electron collisional data used.  It is also worth to note here that the predicted centre-to-limb variation is also practically unchanged in the dwarf models, and thus the same is expected in the Sun.  

The giant atmospheric models, however, show significant differences in $\Delta \rm{A(Mg)}$.  This result is not surprising, as the 457~nm line in giants was singled out in Table~1 of \citeauthor{osorio_mg_2015} as being sensitive to the electron collision data, and we have already noted the significant differences between the M88 and CCC/BSR data for this transition.  The low densities in the line forming regions of giant atmospheres means this difference has a large impact on the statistical equilibrium.  For the solar metallicity case, abundance corrections are $\sim$0.1 dex larger in the M88 model than in both the CCC and BSR models, while for the metal-poor case, the difference in $\Delta \rm{A(Mg)}$ between M88 and BSR/CCC is $\sim$0.08 dex. This difference stems from the much larger collision rate for this transition in the new datasets, mentioned in the previous section.  The abundance corrections obtained in \citeauthor{osorio_mg_2015} are also shown in Fig \ref{fig:acorr} for comparison (RMPS), and also seen to be larger than those using BSR or CCC.  This stems from the fact that the rate for this transition at temperatures of order 5000~K is roughly a factor of two smaller in the RMPS calculations of \citeauthor{osorio_mg_2015} than in the CCC and BSR datasets.  Thus, the new data provide an important improvement to the results of \cite{osorio_mg_2015} and  \cite{osorio_mg_2016} for this line in giant stars.  We note that the oscillator strength for this line has recently been calculated by \cite{rhodin_experimental_2017}, with $\log gf$ increasing by more than 0.2~dex (60~\%) compared to the value from NIST \citep{NIST_5.4} used in those works.  A broader and more detailed study including both of these improvements, and comparison with standard stars should be performed, and this will be the subject of future work.

\subsection{Supernovae ejecta}

Accurate thermal collision strengths are critical for modelling of the
\ion{Mg}{i} 457~nm intercombination line in supernovae ejecta in the emission line phase ($t\ga 100$ days post explosion). Although the neutral fraction is quite low in typical models,
$\sim 10^{-3}$ for Mg, the 457~nm line can still be an important coolant. The electron
densities are generally below the critical density, so the line is formed in non-LTE,
with a strength directly proportional to the collision strength.

\begin{figure}
\includegraphics[width=0.5\textwidth]{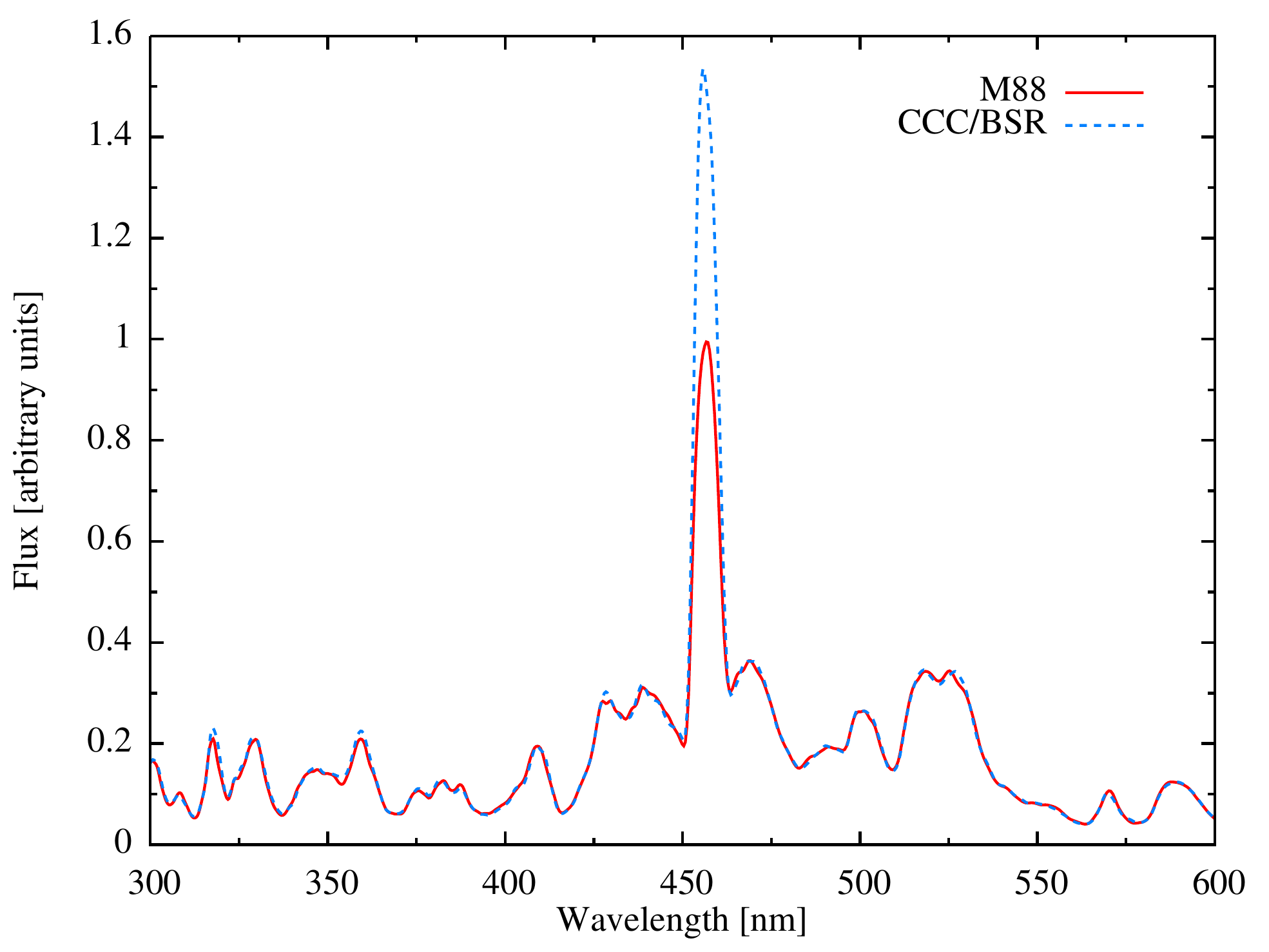}
\caption{Model spectra for a supernova from a $M_\mathrm{ZAMS}=13 M_\sun$
star showing the \ion{Mg}{i} 457~nm line computed with electron collisions data from M88 (full red line) and with the CCC/BSR data from this work (dashed blue line).}\label{fig:sn}
\end{figure}

Figure \ref{fig:sn} compares model spectra of a stripped-envelope supernova from a $M_\mathrm{ZAMS}=13 M_\sun$
star \citep{jerkstrand_late-time_2015}, computed with the M88 collision strengths and
the values calculated here. At the Mg zone temperature in the model (T$\sim$3200K),
the CCC/BSR collision strengths calculated here are about a factor two higher than in M88. This translates in this model to
a factor 1.5 brighter \ion{Mg}{i} 457~nm line, as there is also some contribution due to
recombination. Depending on the model and epoch, the line can change by both
more or less, sometimes up to the full factor of two difference.  
This model improvement translates directly to new (lower) estimates of
the magnesium mass in the ejecta.  It alleviates the discrepancy where the 457~nm line in typical models with Mg/O ratios close to solar tended to be too weak compared to observations \citep{jerkstrand_late-time_2015}.

\section{Conclusions}

The agreement between the CCC and BSR results suggests that the 
effective collision strengths from these methods, and thus the rate coefficients, 
are reliable in most cases to better than 30\%.  Comparison with the data 
from M88 suggests those data are only reliable within a factor of 6 or so.
Thus, the modern close-coupling results are to be strongly 
preferred for non-LTE modelling.  In particular, it is found that the collision strength for the transition corresponding to the \ion{Mg}{i} 457~nm line is significantly underestimated.  The improved data provided here leads to important changes in predictions of the \ion{Mg}{i} 457~nm line in non-LTE line formation models of late-type stellar atmospheres and for supernovae ejecta.

\begin{acknowledgements}P.S.B.\ gratefully acknowledges the support of the 
Swedish Research Council.  P.S.B.\ is also supported by the project grant ``The New Milky Way'' from 
the Knut and Alice Wallenberg foundation.  Y.O. acknowledges funds from the Alexander von Humboldt Foundation in the framework of the Sofja Kovalevskaja Award endowed by the Federal Ministry of Education and Research.  D.F. and I.B. Acknowledge the support of the Australian Research Council, the National Computer Infrastructure, and the Pawsey Computer Centre of Western Australia.  O.Z.\ and K.B.\ acknowledge NSF support through grants 
No.~PHY-1403245, No.~PHY-1520970, and the XSEDE super\-computer allocation No.~PHY-090031.
A.J.\ acknowledges funding by the European Union’s Framework Programme for Research and Innovation Horizon 2020 under Marie Sklodowska-Curie grant agreement No 702538.
\end{acknowledgements}

\bibliographystyle{aa} 
\bibliography{../../../../MyLibrary.bib}

\end{document}